\documentclass[preprint2]{aastex}

\shorttitle{SNe Ia and SNe II contributions to metals in the ICM}
\shortauthors{Sato et al.}

\begin{document}


\title{Type Ia and II supernovae contributions 
to the metal enrichment \\
in intra-cluster medium observed with {\it Suzaku}}


\author{
 Kosuke \textsc{Sato},\altaffilmark{1,2}
 Kazuyo \textsc{Tokoi},\altaffilmark{2}
 Kyoko \textsc{Matsushita},\altaffilmark{1} 
 Yoshitaka \textsc{Ishisaki},\altaffilmark{2} \\ 
 Noriko \textsc{Y.~Yamasaki},\altaffilmark{3}
 Manabu \textsc{Ishida},\altaffilmark{3} 
 \textsc{and} Takaya \textsc{Ohashi}\,\altaffilmark{2} 
}


\altaffiltext{1}{
Department of Physics, Tokyo University of Science,
1-3 Kagurazaka, Shinjuku-ku, Tokyo 162-8601, Japan: ksato@rs.kagu.tus.ac.jp}
\altaffiltext{2}{
Department of Physics, Tokyo Metropolitan University,
 1-1 Minami-Osawa, Hachioji, Tokyo 192-0397, Japan}
\altaffiltext{3}{
Institute of Space and Astronautical Science (ISAS),
Japan Aerospace Exploration Agency, 
3-1-1 Yoshinodai, Sagamihara, Kanagawa 229-8510, Japan}


\begin{abstract}

We studied the properties of the intra-cluster medium (ICM) in
two clusters of galaxies (AWM~7 and Abell~1060) and
two groups (HCG~62 and NGC~507) with the X-ray Observatory {\it Suzaku}.
Based on spatially resolved energy spectra, we measured for the first time
precise cumulative ICM metal masses within 0.1 and $\sim 0.3\; r_{180}$.
Comparing our results with supernova nucleosynthesis models,
the number ratio of type II (SNe~II) to type Ia (SNe~Ia)
is estimated to be $\sim 3.5$, assuming the metal mass in the ICM
is represented by the sum of products synthesized in SNe~Ia and SNe~II\@.
Normalized by the K-band luminosities of present galaxies,
and including the metals in stars, the integrated number of past
SNe~II explosions is estimated to be close to or somewhat higher than
the star formation rate determined from Hubble Deep Field observations.

\end{abstract}



\section{Introduction}

The elemental abundance of the intra-cluster medium (ICM) gives us
important clues for understanding the chemical history and evolution
process of clusters of galaxies.  Because clusters are the largest
gravitationally bound systems in the universe, they are expected to
confine all the metals provided by member galaxies since the cluster
formed.  The amount of each metal in the ICM is almost equal to the
integrated sum of that synthesized by supernovae type Ia (SNe~Ia) and
type II (SNe~II)\@.  The expected elemental products of each type have
been intensively studied by several authors \citep{iwamoto99,nomoto06}.
X-ray measurements of the mass of each element in the ICM enable us to
examine the past occurrence numbers and the ratio of SNe~II to SNe~Ia.
These numbers also reflect the past stellar initial mass function
(IMF) and star formation rate (SFR) in clusters.

{\it ASCA} first measured metal distributions in the ICM 
\citep{mushotzky96,fukazawa98,fukazawa00}, and
\citet{dupke00} and \citet{baumgartner05} investigated 
the contributions from SNe Ia and SNe~II to the ICM\@.
Recent X-ray observations with {\it XMM-Newton} determined the
spatial distribution and elemental abundance pattern of the ICM metals
\citep{matsushita03,matsushita07b,tamura04,boehringer05,deplaa06,deplaa07,
werner06}.
However, the abundance profiles in the outer regions of clusters 
are still poorly constrained,
especially for O and Mg which are major SNe~II products,
because of the intense intrinsic Al-K backgrounds
and the non-Gaussian line profiles of previous missions in the energy 
range below 1 keV\@.
Based not only on the low and stable background
but also on the good sensitivity to emission lines below $\sim 1$~keV
of the {\it Suzaku} XIS \citep{koyama07},
a reliable determination of O and Mg abundances to cluster outer regions
has become feasible \citep{matsushita07a,sato07}.

We use $H_0=70$ km s$^{-1}$ Mpc$^{-1}$ = $0.7\;h_{\rm 100}$,
$\Omega_{\rm \Lambda}= 1-\Omega_M=0.73$,
and the virial radius is assumed
$r_{\rm 180} = 1.95\; h_{100}^{-1}\sqrt{k\langle T\rangle/10~{\rm keV}}$~Mpc
\citep{markevitch98} in this paper.
Errors are 90\% confidence region for a single
interesting parameter.

\begin{table*}[tb]
\caption{
Integrated number of SNe~I ($N_{\rm Ia}$) and
number ratio of SNe~II to SNe~Ia ($N_{\rm II}$/$N_{\rm Ia}$).
}\label{tab:1}
\begin{center}
\begin{tabular}{lllcccr}
\tableline\tableline
\makebox[8em][l]{Object} &
\multicolumn{1}{r}{\makebox[0in][r]{$z$ / $k\langle T\rangle$ / $r_{180}$}} &
\multicolumn{1}{c}{Region} &
\makebox[0in][c]{SNe~Ia Model} &
$N_{\rm Ia}$& $N_{\rm II}$/$N_{\rm Ia}$ & $\chi^2$/dof \\
\tableline
AWM~7 $\dotfill$ & 0.01724 &
 $<0.1\;r_{180}$ &W7 & $1.2\pm0.2\times10^{9}$& $4.0\pm 1.2$ & 12.6/3\\
& 3.5 keV &
 $< 0.35\;r_{\rm 180}$ &W7 & $7.6\pm1.0\times10^{9}$& $3.7\pm 1.2$ & 15.9/3\\
& 1.65 Mpc &
 $< 0.35\;r_{\rm 180}$ &WDD1 & $9.0\pm1.2\times10^{9}$& $2.3\pm 1.1$ & 35.5/3\\
&  &
 $< 0.35\;r_{\rm 180}$ &WDD2 & $7.3\pm1.0\times10^{9}$& $3.5\pm 1.3$ & 20.5/3\\
\tableline
A~1060 $\dotfill$ & 0.0114 &
 $<0.1\;r_{180}$ &W7 & $5.1\pm1.1\times10^{8}$& $3.0\pm 1.6$ & 4.5/3\\
& 3.0 keV &
 $<0.25\;r_{\rm 180}$ &W7& $1.5\pm0.3\times10^{9}$& $2.6\pm 1.5$ & 8.5/3\\
& 1.53 Mpc &
 $<0.25\;r_{\rm 180}$ &WDD1& $1.6\pm0.4\times10^{9}$& $2.0\pm 1.4$ & 16.0/3\\
&  &
 $<0.25\;r_{\rm 180}$ &WDD2& $1.4\pm0.3\times10^{9}$& $2.6\pm 1.6$ & 10.2/3\\
\tableline
NGC~507 $\dotfill$ & 0.01646 &
 $<0.1\;r_{180}$ & W7 & $8.2\pm1.7\times10^{7}$& $4.2\pm 1.4$ & 6.6/3\\
& 1.5 keV &
 $<0.24\;r_{\rm 180}$ &W7 & $3.1\pm0.6\times10^{8}$& $3.7\pm 1.2$ & 11.7/3\\
& 1.08 Mpc &
 $<0.24\;r_{\rm 180}$ &WDD1 & $3.3\pm0.7\times10^{8}$& $2.9\pm 1.2$ & 22.0/3\\
&     &
 $<0.24\;r_{\rm 180}$ &WDD2 & $3.0\pm0.6\times10^{8}$& $3.6\pm 1.2$ & 13.1/3\\
\tableline
HCG~62 $\dotfill$ &0.0145 &
 $<0.1\;r_{180}$ &W7 &$1.0\pm0.2\times10^{8}$& $3.1\pm 1.3$ & 4.2/3 \\
 & 1.5 keV &
 $< 0.21\;r_{\rm 180}$ &W7 & $2.8\pm0.6\times10^{8}$ & $2.6\pm 1.0$ & 3.0/3 \\
 & 1.08 Mpc &
 $< 0.21\;r_{\rm 180}$ &WDD1 & $2.7\pm0.6\times10^{8}$ &$2.0\pm1.1$ & 12.7/3 \\
 &  &
 $< 0.21\;r_{\rm 180}$ &WDD2 & $2.6\pm0.5\times10^{8}$ &$2.5\pm1.1$ & 4.8/3 \\
\tableline
\end{tabular}
\end{center}
\end{table*}

\begin{figure*}[t]
\epsscale{1.0}
\plotone{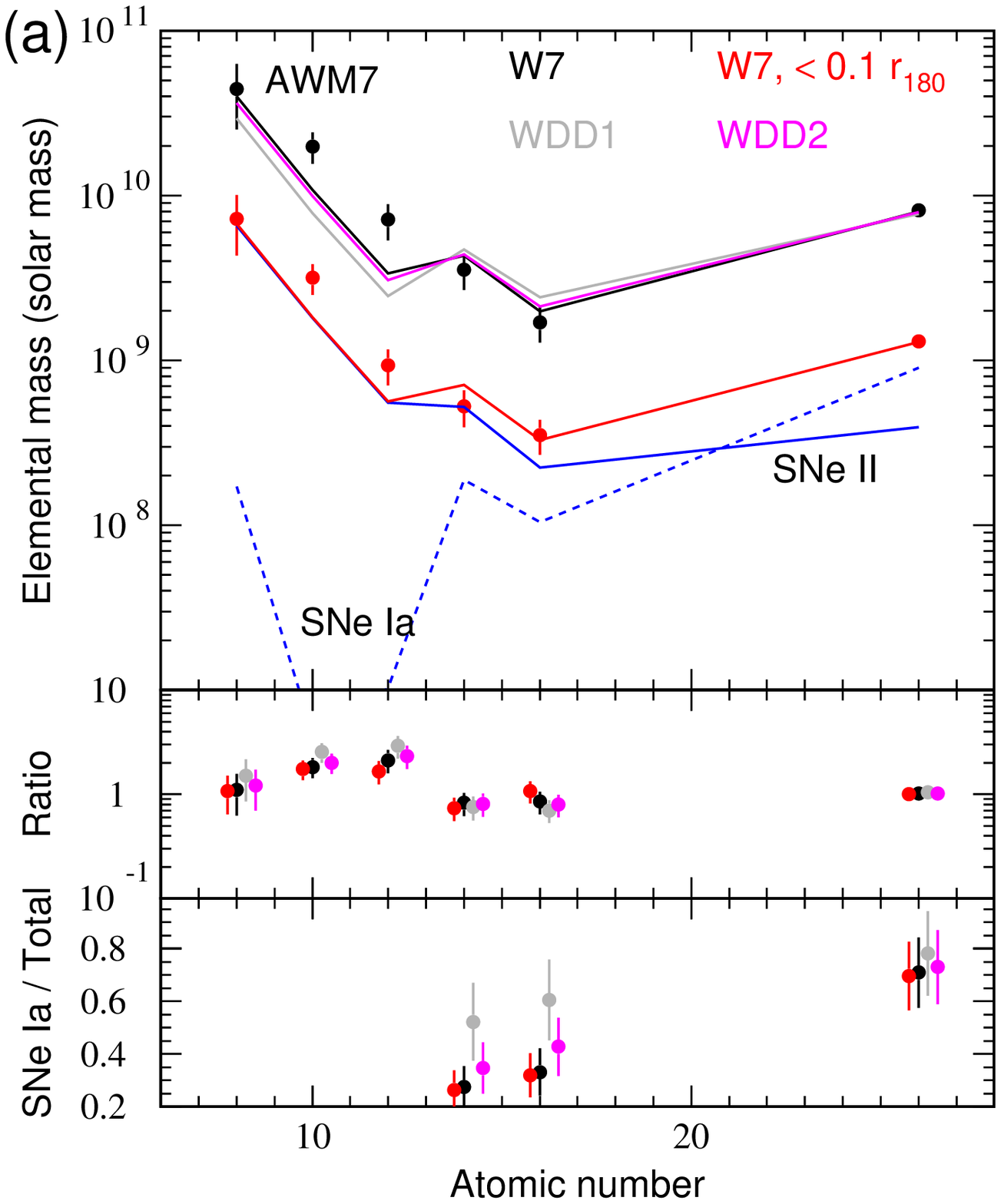}
\hfill
\plotone{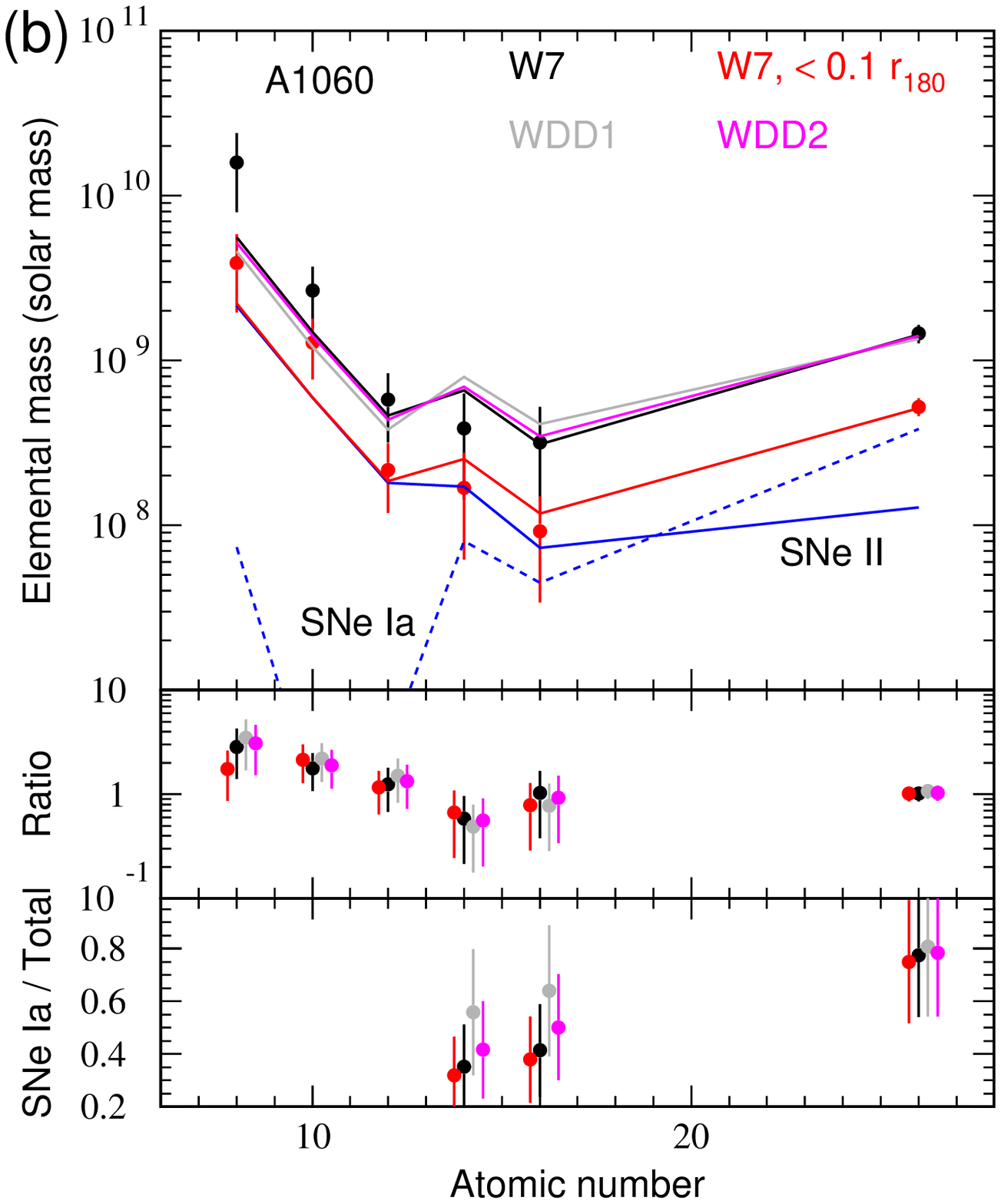}
\plotone{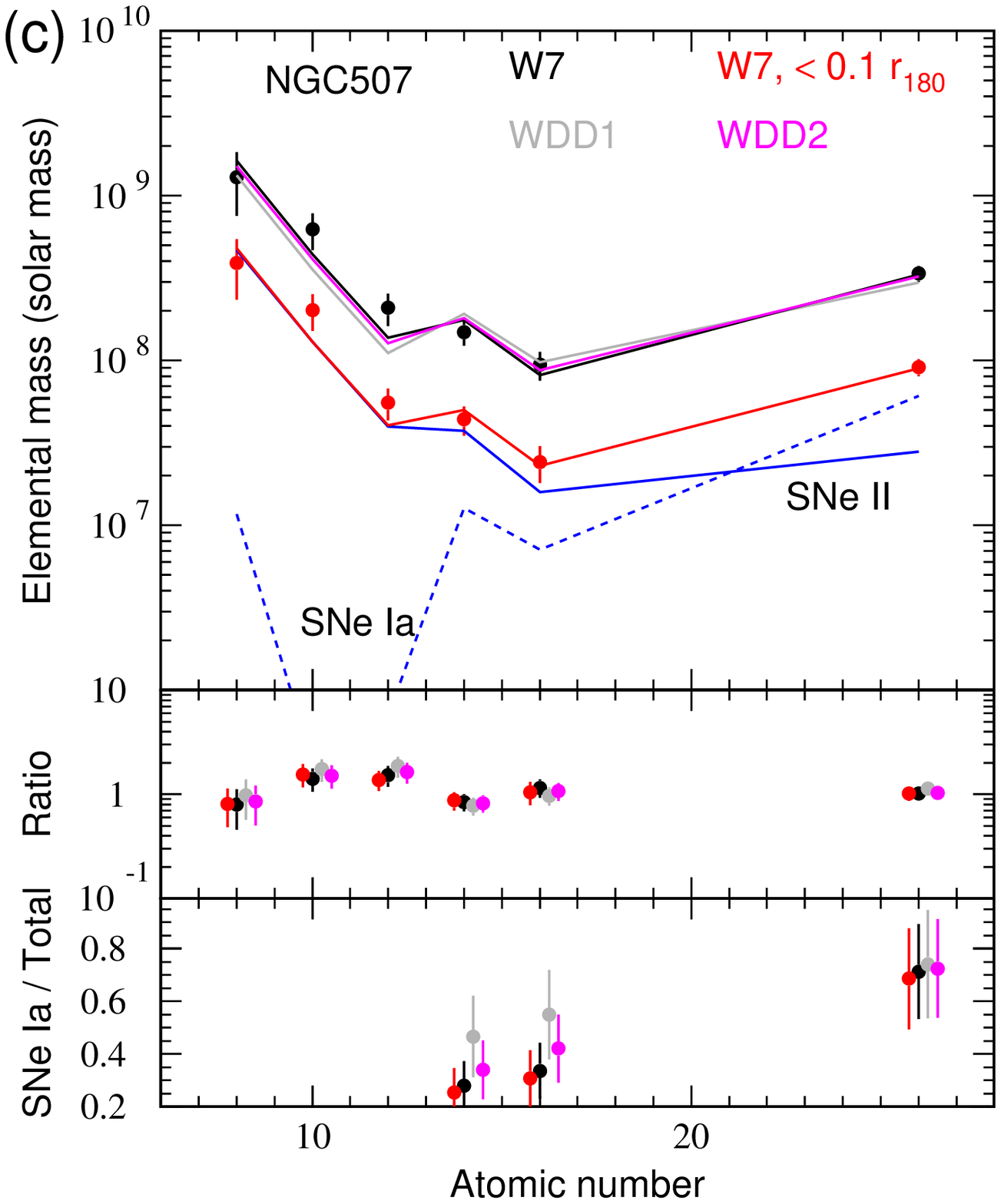}
\hfill
\plotone{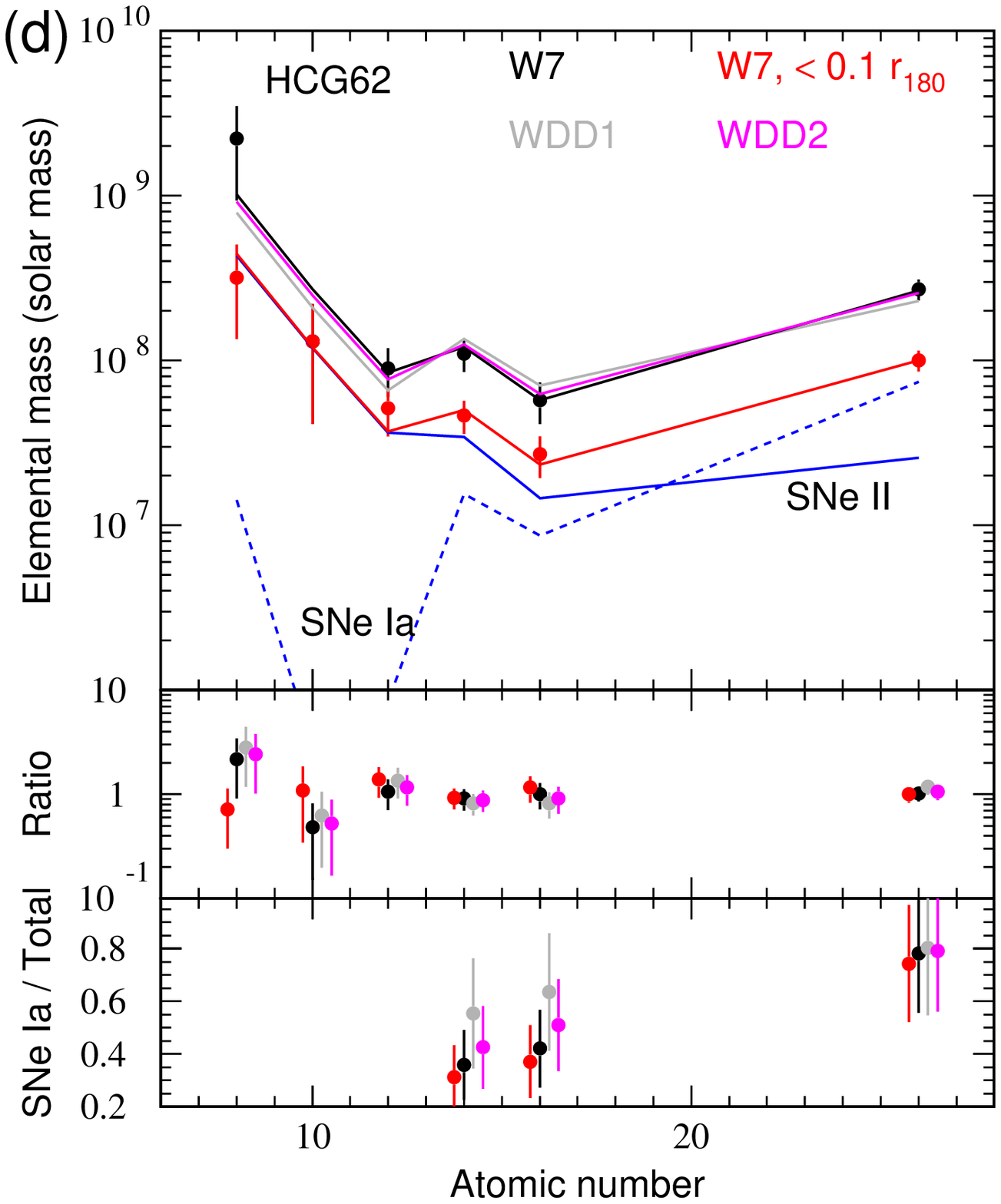}
\caption{
Fit results of each elemental mass for
(a)~AWM~7, (b)~A~1060, (c)~NGC~507, and (d)~HCG~62\@.
Top panels show the mass within the whole observed region (black)
and within $0.1\;r_{\rm 180}$ (red) fit by
$\left[\, N_{\rm Ia}\; \{ ({\rm SNe~Ia\ yield}) +
(N_{\rm II}/N_{\rm Ia})\; ({\rm SNe~II\ yield})\}\,\right]$.
Blue dashed and solid lines corresponds to the contributions 
of SNe~Ia (W7) and SNe~II within $0.1\;r_{180}$, respectively.
Ne (atomic number = 10) is excluded in the fit.
Mid and lower panels indicate ratios of data points to the best-fit,
and fractions of the SNe~Ia contribution to total mass in the best-fit
model for each element, respectively.
}\label{fig:1}
\end{figure*}

\section{Metal mass determination}

We selected two clusters, AWM~7 and Abell~1060 (hereafter A~1060), and
two groups, HCG~62 and NGC~507, that were observed with {\it Suzaku}.
Detailed information and the analysis method for each object are
described in \citet{sato07,sato07b,sato07c} and \citet{tokoi07}.  We
excluded Ni and Ne abundances in our analysis in this paper since
these two elements were not reliably determined due to the strong and
complex Fe-L line emissions.  For the first time, we determined the
radial abundance profiles of O, Mg, Si, S and Fe, out to
$\sim0.3~r_{180}$ for each object.  Note that the O abundance in the
ICM is strongly affected by the foreground Galactic emission, because
we cannot resolve the cluster and local lines.  Combining the
abundance profile obtained with {\it Suzaku} and the X-ray luminous
gas mass profile with {\it XMM-Newton}, we calculated cumulative metal
mass within $0.1~r_{180}$ and the whole observed region
(table~\ref{tab:1}).  Errors of the metal mass plotted in
figure~\ref{fig:1} are taken from the statistical errors of each
elemental abundance in the spectral fits at $\sim0.1~r_{180}$.

\section{Deconvolution of elemental mass patterns with SNe yields}

\subsection{Mass products by SNe~Ia and II}

\begin{figure*}[tb]
\epsscale{1.0}
\plotone{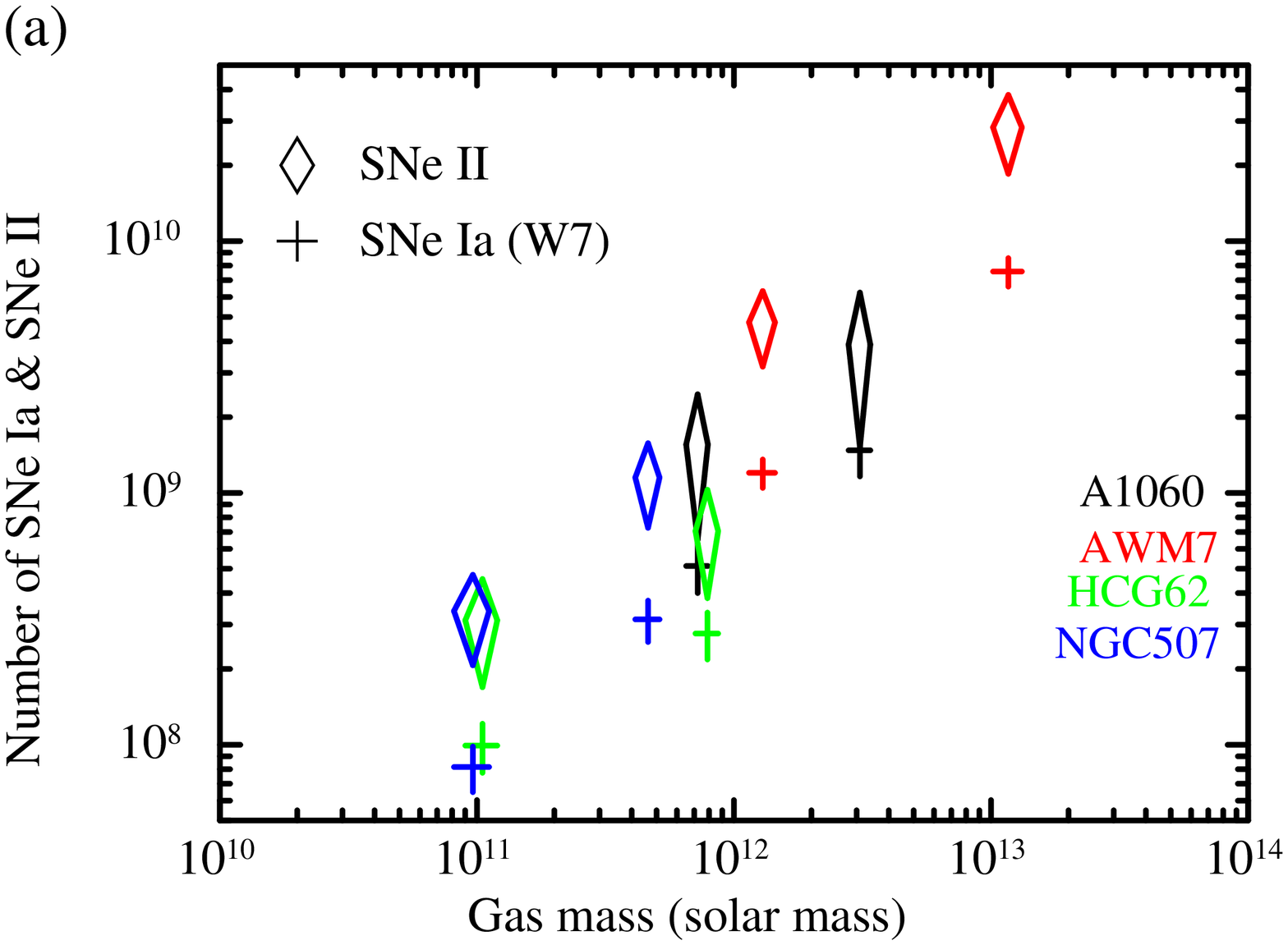}
\hfill
\plotone{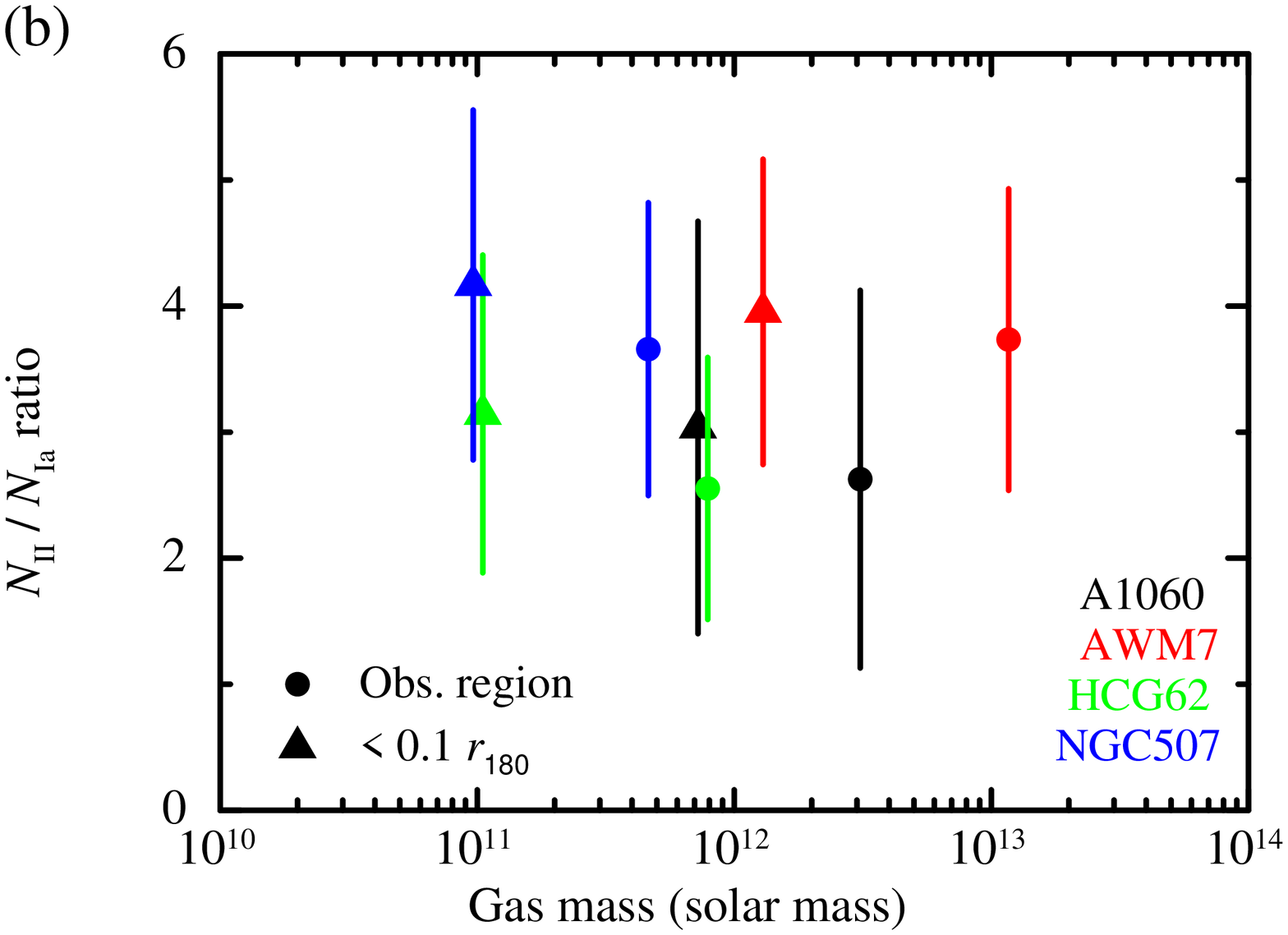}
\caption{
(a) Integrated numbers of SNe~Ia (W7) and SNe~II
plotted against gas mass within the calculated region.
(b) Number ratio of SNe~II to SNe~Ia with W7 model.
\label{fig:2}}
\end{figure*}

In order to examine the SNe~Ia and SNe~II contribution to the ICM
metals, the elemental mass pattern of O, Mg, Si, S and Fe was examined
within $0.1\;r_{180}$ and the whole observed region. The mass patterns
were fit by a combination of average SNe~Ia and SNe~II yields per
supernova, as shown in figure~\ref{fig:1}.  The fit parameters were
chosen to be the integrated number of SNe~Ia ($N_{\rm Ia}$) and the
number ratio of SNe~II to SNe~Ia ($N_{\rm II}/N_{\rm Ia}$), because
$N_{\rm Ia}$ could be well constrained due to relatively small errors
in the Fe abundance.  The SNe Ia and II yields were taken from
\citet{iwamoto99} and \citet{nomoto06}, respectively.  We assumed a
Salpeter IMF for stellar masses from 10 to 50 $M_{\odot}$ with the
progenitor metallicity of $Z=0.02$ for SNe~II, and W7, WDD1 or WDD2
models for SNe~Ia.  Table~\ref{tab:1} and figure~\ref{fig:2}
summarizes the fit results.

All the objects exhibit similar features.  The abundance patterns were
better represented by the W7 SNe~Ia yield model rather than WDD1.  The
number ratio of SNe~II to SNe~Ia with W7 is $\sim 3.5$, while the
ratio with WDD1 is $\sim 2.5$.  The WDD2 model gave quite similar
results as W7\@.  Almost 3/4 of the Fe and $\sim 1/4$ of the Si is
synthesized by SNe~Ia, in the W7 model, as demonstrated in lower
panels of figure~\ref{fig:1}.  We also examined the SNe~II yields with
$Z=0.004$ in \citet{nomoto06}, and the results tended to show smaller
$N_{\rm II}/N_{\rm Ia}$ ratios by $\sim30$\% and larger $\chi^2$
values.  Somewhat earlier estimates of SNe~II yields summarized in
\citet{iwamoto99} are between $Z=0.02$ and $Z=0.004$, but our $\chi^2$
values are almost the same with those of $Z=0.02$.  The values in
Table 1 imply that the fit to the data and the $N_{\rm II}/N_{\rm Ia}$
ratio behave in a similar manner for different supernova models
between $r < 0.1\;r_{180}$ and $r > 0.1\;r_{180}$.

We note that most of the fits were not formally acceptable based on
the $\chi^2$ values in table~\ref{tab:1}.  The 90\% and 99\%
confidence values for 3 degrees of freedom are $\chi^2=6.25$ and
11.34, respectively.  The models adapted here are probably too
simplified, such as assumptions of uniform IMF and SNe yields at all
times, spherical symmetry in the ICM, and rather crude treatment of
the Galactic emission, etc.  For example, a flatter slope of the IMF
than Salpeter tends to yield a larger amount of O\@.  It is notable
that the observed amount of Mg is about twice as much as the model
prediction for AWM7 and NGC~507, similar to the feature reported by
\citet{werner06b} for M~87\@.

\subsection{Comparison with other studies}
\label{subsec:comparison}

\citet{deplaa07} also found that the number ratio of core-collapse SNe
(SNe~II+Ib+Ic) to SNe~Ia (using W7) is $\sim 3.5$, based on {\it
XMM-Newton} observations of 22 clusters within $0.2\;r_{500}$ for
elements between Si and Ni.  They also examined the fit including the
average O and Ne abundances for two clusters observed with the XMM
Reflection Grating Spectrometers (RGS), however RGS observations are
limited to the central cluster region where the abundance is strongly
affected by the cD galaxy.  The number ratio of SNe~II to SNe~Ia they
have obtained is consistent with our results.

\citet{tsujimoto95} determined that the number ratio of SNe~II to
SNe~Ia for our Galaxy is $\sim 6.7$ based on stellar metallicity,
while it is 3.3--5 for the Large and Small Magellanic Clouds.  These
results are slightly higher than those for our clusters and groups.
It is plausible that spiral galaxies in the Local Group maintain star
forming activity until recently.  We also note that our results are
derived for metals only in the ICM, therefore metals in member
galaxies (including stars) which accounts for $\sim 50$\% of the total
are not considered.

\subsection{SNe~II per unit galaxy luminosity}

\begin{table}[tb]
\begin{center}
\caption{
Number of SNe~II ($<0.1~r_{180}$) per volume ($N_{\rm II}/V$),
K-band galaxy luminosity per volume ($L_{\rm K}/V$),
and $N_{\rm II}/L_{\rm K}$ in HDF and clusters.
}\label{tab:3}
\begin{tabular}{lclc}
\tableline\tableline
 & $N_{\rm II}/V$ & \multicolumn{1}{c}{$L_{\rm K}/V$} & $N_{\rm II}/L_{\rm K}$ \\
& (Mpc$^{-3}$) & ($L_\odot$~Mpc$^{-3}$) & ($L_\odot^{-1}$) \\
\tableline
HDF & $\lesssim 10^{5}$ & \ \ 4.0$\times 10^{8}$ & 2.5$\times 10^{-3}$\\
AWM~7$^\ast$ & $2.2\times 10^{11}$& \ \ 5.4$\times 10^{13}$ &4.1$\times 10^{-3}$\\
A~1060$^\ast$ & $8.2\times 10^{10}$& \ \ 5.1$\times 10^{13}$& 1.6$\times 10^{-3}$\\
\tableline
\multicolumn{4}{l}{\parbox{0.45\textwidth}{\footnotesize 
$^\ast$
For AWM~7 and A~1060, the number of SNe~II is estimated from 
the metals of the ICM, not including the member galaxies and stars.}}\\
\end{tabular}
\end{center}
\end{table}

Based on the luminosity function at high redshift, the star formation
history of the universe has been studied.  \citet{madau98} derived a
simple stellar evolution model for field galaxies in the Hubble Deep
Field (HDF)\@.  The SFR rises sharply, by about an order of magnitude,
from the present to a peak value in the range 0.12--0.17 $M_{\odot}$
yr$^{-1}$~Mpc$^{-3}$ at $z\sim 1.5$.  \citet{mannucci07} and
\citet{nagamine06} also derived the cosmic star formation history from
various observations, with no correction for dust extinction in
cosmology of $h_{100}=0.7$ and $\Omega_{\rm \Lambda}=0.7$.

If we assume the maximum SFR is 0.1 $M_{\odot}$ yr$^{-1}$~Mpc$^{-1}$
and integrate it over a Hubble time, the total produced stellar
mass is $10^{9}\;M_{\odot}$~Mpc$^{-1}$.
Supposing a Salpeter IMF between 10--50 $M_{\odot}$,
estimated number of SNe~II per unit volume
is $\lesssim 10^5$ Mpc$^{-3}$ for field galaxies.
When we scale our cluster results only by volume, the
normalized numbers of SNe~II within 0.1 $r_{180}$ 
for AWM~7 and A~1060 are $2.2\times 10^{11}$ and
$8.2\times 10^{10}$ Mpc$^{-3}$, respectively.
These factors of $\sim10^{5}$ discrepancies are primarily caused
by  the difference in the number of galaxies between the field and clusters.
We therefore scale these results with the K-band luminosity density
as summarized in table~\ref{tab:3}.
Using an average luminosity density
$L_{\rm K}/V = (5.74\pm 0.86)\times 10^{8}\;h_{100}\;L_{\odot}$ Mpc$^{-3}$ 
in \citet{cole05}, the normalized number of SNe~II 
per unit galaxy luminosity is
$2.5\times 10^{-3}\;L_{\odot}^{-1}$ for field galaxies.
The galaxy luminosity densities of AWM~7 and A~1060
are taken from the Two Micron All Sky Survey.\footnote{
The database address: {\tt http://www.ipac.caltech.edu/2mass/}}
As shown in table~\ref{tab:3}, the
numbers of SNe~II per unit luminosity in the clusters are
quite consistent with the above value independently estimated
from the star formation history.
However, this estimate considers only the ICM metals, and
inclusion of stellar metals would roughly double the required
SNe II rate. 
Even though systematic uncertainty in this simple
estimation may account for most of the factor of two difference, it is
interesting that X-ray metals suggest incompleteness in the optical
detection of distant star formation. Also, an environmental
difference between fields and clusters may be of relevance.

\acknowledgments
Authors are grateful to  N.~Arimoto, K.~Masai, S.~Sasaki, and P. Henry 
for valuable comments and discussions.
We also thank the referee for providing valuable comments.
Part of this work was financially supported by the Ministry of
Education, Culture, Sports, Science and Technology of Japan,
Grant-in-Aid for Scientific Research
No.\ 14079103, 15340088, 15001002, 16340077, 18740011.



\begin{thebibliography}{}

\bibitem[Baumgartner et al.(2005)]{baumgartner05} Baumgartner, W.~H., 
Loewenstein, M., Horner, D.~J., \& Mushotzky, R.~F.\ 2005, \apj, 620, 680 

\bibitem[B{\"o}hringer et al.(2005)]{boehringer05} B{\"o}hringer, 
H., Matsushita, K., Finoguenov, A., Xue, Y., \& Churazov, E.\ 2005, 
Advances in Space Research, 36, 677 

\bibitem[Cole et al.(2005)]{cole05} Cole, S., et al.\ 2005,
\mnras, 362, 505

\bibitem[de Plaa et al.(2007)]{deplaa07} de Plaa, J., Werner,
N., Bleeker, J.~A.~M., Vink, J., Kaastra, J.~S., \& M{\'e}ndez, M.\ 2007,
\aap, 465, 345

\bibitem[de Plaa et al.(2006)]{deplaa06} de Plaa, J., et al.\
2006, \aap, 452, 397

\bibitem[Dupke \& White(2000)]{dupke00} Dupke, R.~A., \& White, 
R.~E., III 2000, \apj, 528, 139 

\bibitem[Fukazawa et al.(2000)]{fukazawa00} Fukazawa, Y., 
Makishima, K., Tamura, T., Nakazawa, K., Ezawa, H., Ikebe, Y., Kikuchi, K., 
\& Ohashi, T.\ 2000, \mnras, 313, 21 

\bibitem[Fukazawa et al.(1998)]{fukazawa98} Fukazawa, Y., 
Makishima, K., Tamura, T., Ezawa, H., Xu, H., Ikebe, Y., Kikuchi, K., \& 
Ohashi, T.\ 1998, \pasj, 50, 187 

\bibitem[Iwamoto et al.(1999)]{iwamoto99} Iwamoto, K., Brachwitz,
F., Nomoto, K., Kishimoto, N., Umeda, H., Hix, W.~R., \& Thielemann, F.-K.\
1999, \apjs, 125, 439

\bibitem[Koyama et al.(2007)]{koyama07}
        Koyama, K., et al.\ 2007, \pasj, 59, 23 

\bibitem[Madau et al.(1998)]{madau98} Madau, P., Pozzetti, L.,
\& Dickinson, M.\ 1998, \apj, 498, 106

\bibitem[Mannucci et al.(2007)]{mannucci07} Mannucci, F., Buttery,
H., Maiolino, R., Marconi, A., \& Pozzetti, L.\ 2007, \aap, 461, 423

\bibitem[Markevitch et al.(1998)]{markevitch98}
        Markevitch, M., et al.\ 1998, \apj, 503, 77

\bibitem[Mushotzky et al.(1996)]{mushotzky96} Mushotzky, R., 
Loewenstein, M., Arnaud, K.~A., Tamura, T., Fukazawa, Y., Matsushita, K., 
Kikuchi, K., \& Hatsukade, I.\ 1996, \apj, 466, 686 

\bibitem[Matsushita et al.(2007a)]{matsushita07a} Matsushita, K., et 
al.\ 2007a, \pasj, 59, 327 

\bibitem[Matsushita et al.(2007b)]{matsushita07b} Matsushita, K., 
B{\"o}hringer, H., Takahashi, I., \& Ikebe, Y.\ 2007b, \aap, 462, 953 

\bibitem[Matsushita et al.(2003)]{matsushita03} Matsushita, K., 
Finoguenov, A., B{\"o}hringer, H.\ 2003, \aap, 401, 443 

\bibitem[Nagamine et al.(2006)]{nagamine06} Nagamine, K.,
Ostriker, J.~P., Fukugita, M., \& Cen, R.\ 2006, \apj, 653, 881

\bibitem[Nomoto et al.(2006)]{nomoto06} Nomoto, K., Tominaga, 
N., Umeda, H., Kobayashi, C., \& Maeda, K.\ 2006, Nuclear Physics A, 777, 
424 

\bibitem[Sato et al.(2007a)]{sato07} Sato, K., et al.\ 2007a,
\pasj, 59, 299

\bibitem[Sato et al.(2007b)]{sato07b} Sato, K., Matsushita, K., 
Ishisaki, Y., Yamasaki, N. Y., Ishida, M. Sasaki, S., \& 
Ohashi, T., \ 2007b, PASJ in press, arXiv:0707.4342

\bibitem[Sato et al.(2007c)]{sato07c} Sato, K., et al.\ 2007c,
in preparation

\bibitem[Tamura et al.(2004)]{tamura04} Tamura, T., Kaastra,
J.~S., den Herder, J.~W.~A., Bleeker, J.~A.~M., \& Peterson, J.~R.\ 2004,
\aap, 420, 135

\bibitem[Tokoi et al.(2007)]{tokoi07} Tokoi, K., et al.\ 2007,
in preparation

\bibitem[Tsujimoto et al.(1995)]{tsujimoto95} Tsujimoto, T.,
Nomoto, K., Yoshii, Y., Hashimoto, M., Yanagida, S., \& Thielemann, F.-K.\
1995, \mnras, 277, 945

\bibitem[Werner et al.(2006)]{werner06} Werner, N., de Plaa, J., 
Kaastra, J.~S., Vink, J., Bleeker, J.~A.~M., Tamura, T., Peterson, J.~R., 
\& Verbunt, F.\ 2006a, \aap, 449, 475 

\bibitem[Werner et al.(2006b)]{werner06b} Werner, N., 
B{\"o}hringer, H., Kaastra, J.~S., de Plaa, J., Simionescu, A., \& Vink, 
J.\ 2006b, \aap, 459, 353 

\end{thebibliography}
\end{document}